\documentclass[12pt,english]{article}
\usepackage[breaklinks,hypertexnames=false]{hyperref}
\hypersetup{colorlinks=true,linkcolor=blue,urlcolor=blue,citecolor=blue}

\usepackage{amsthm}
\usepackage{amsmath}
\usepackage{amssymb}
\usepackage{esint}
\usepackage[authoryear,longnamesfirst,round]{natbib}
\usepackage{pgfplots}
\pgfplotsset{compat=1.18}

\usepackage{subcaption} 

\usepackage{appendix}
\usepackage[english]{babel}
\usepackage{enumitem}
\usepackage{booktabs}
\usepackage{multirow}
\usepackage{natbib}
\usepackage{textcomp,mathcomp}
\usepackage{adjustbox}
\usepackage{dcolumn}
\usepackage{booktabs}
\usepackage{mathabx} 
\usepackage{tcolorbox} 
\usepackage{enumitem}  

\makeatletter
\usepackage{graphicx}
\usepackage{lscape}
\usepackage{rotating}
\usepackage[english]{babel}
\usepackage{latexsym}	
\usepackage{bbm}
\usepackage{tikz}
\usetikzlibrary{arrows,shapes,trees,positioning,decorations.markings,shapes.geometric, arrows.meta}

\usepackage{seqsplit}
\usepackage{graphicx}
\hypersetup{
	colorlinks=true,
	citecolor=blue
}

\makeatother

\newtheorem*{example*}{Example}

\newcommand{\defeq}{\vcentcolon=}

\usepackage{todonotes}
\usepackage{tikz}

\newcommand{\wheelchart}[3]{
	\pgfmathsetmacro{\totalnum}{0}
	\foreach \value/\colour in {#1} {
		\pgfmathparse{\value+\totalnum}
		\global\let\totalnum=\pgfmathresult
	}
	
	\pgfmathsetmacro{\wheelwidth}{(#3)-(#2)}
	\pgfmathsetmacro{\midradius}{(#3+#2)/2}
	
	\begin{scope}[rotate=90]
		\pgfmathsetmacro{\cumnum}{0}
		\foreach \value/\colour in {#1} {
			\pgfmathsetmacro{\newcumnum}{\cumnum + \value/\totalnum*360}
			
			\draw[fill=\colour] (-\cumnum:#2) arc (-\cumnum:-\newcumnum:#2)--(-\newcumnum:#3) arc (-\newcumnum:-\cumnum:#3)--cycle;
			
			\global\let\cumnum=\newcumnum
		}
	\end{scope}
}

\input{tcilatex}
\usepackage{babel}

\begin{document}
	
		\title{{\LARGE {A Users' Guide to Uncovering Worker and Firm Effects:\\The ABC of AKM}\thanks{The authors acknowledge helpful comments by Martyn Andrews, Erik Hurst, Jeffrey Kling, Timothy Taylor, and Heidi Williams. St\'ephane Bonhomme is Professor of Economics and Thibaut Lamadon is Associate Professor of
Economics, both at the University of Chicago, Chicago, Illinois. Elena Manresa is Professor of
Economics, Princeton University, Princeton, New Jersey. Their email addresses are
sbonhomme@uchicago.edu, lamadon@uchicago.edu, and em7118@princeton.edu.}\\$\quad$\\$\quad$}}

	\date{\today}
		
	\maketitle
	\vskip 1cm
	
	\begin{abstract}
		
		The AKM model introduced by \citet{abowd1999high} has become a workhorse to study worker and firm heterogeneity, and to understand the sources of wage dispersion in the labor market using linked employer-employee data. In this article, we introduce the model and estimator, discuss some best practices for estimation, and review some empirical findings on the role of worker and firm heterogeneity in wage dispersion. While the AKM methodology has proven useful to analyze a host of questions in a variety of settings within labor economics and beyond, we also point to the need for methodological developments.

		\bigskip
		
		
	\end{abstract}
	
	\clearpage
	
	\global\long\def\ind{\mathbb{1}}
	\global\long\def\d{\mathrm{d}}
	\global\long\def\t{\intercal}
	\global\long\def\RR{\mathbb{R}}
	\global\long\def\defeq{:=}

\vskip .3cm \noindent Linked employer-employee data are increasingly used to study key questions about labor markets, related to worker and firm heterogeneity, sorting, and the sources of wage dispersion. The main feature of linked data is that they keep track of the identity of workers and firms while following workers across employers. While initially rare and difficult to access for confidentiality reasons, linked data sets have become inceasingly widely available and have been used to study a host of economic questions.

\vskip .3cm \noindent The workhorse method in the literature is the so-called AKM method introduced in \citet*{abowd1999high}. The insight of AKM is to use linked data in order to tell apart the roles of worker and firm heterogeneity in wage determination. By estimating firm and worker components, researchers can then answer a variety of questions related to the wage potentials of workers, the pay policies of firms, and the sorting patterns between workers and firms. 

\vskip .3cm \noindent A key motivation for the method is to identify the sources of the wage differences across firms that characterize modern labor markets: inter-industry differentials are sizable, larger firms tend to pay better than smaller ones, multinationals pay higher wages than national firms, and there are also substantial within-group wage differentials (say, between large firms in a similar industry). However, these differences  may come from different sources: they may arise because firms pay similar workers differently, or because they employ different types of workers. The first explanation points to the existence of high- and low-paying firms, while the second one points to high- and low-wage workers being employed by different firms. 


\vskip .3cm \noindent The AKM methodology provides a way to quantify these two mechanisms, and to document how heterogeneity in permanent differences among workers, and heterogeneity in permanent differences in firms' pay policies, shape individual wages. The AKM model postulates that, in addition to worker and firm characteristics that are observed in the data (such as the worker's experience or the firm's size), the wage is determined by two key factors. The first component, denoted as $\psi_j$, is specific to the firm. Hence, two firms may pay the same worker differently, implying the existence of firm-specific wage premia. The second component, denoted as $\alpha_i$, is specific to the worker. Hence, two workers in the same firm may earn different wages. The AKM model postulates that, net of covariates and some idiosyncratic shocks, wages (expressed in logarithms) are additive in the worker-specific and firm-specific components.

\begin{figure}
\begin{center}
\caption{Wages over time in the AKM model\label{fig_eventstudy}}\definecolor{myBlue}{HTML}{483D8B}
\definecolor{myGray}{HTML}{696969}
\definecolor{myRed}{HTML}{8B0000}

\begin{tikzpicture}
\begin{axis}[xlabel={Period ($t$)},
    ylabel={Log wage ($\alpha_i+\psi_j$) },
    xmin=0.5, xmax=6.7,
    ymin=0, ymax=4.5,
    xtick={0,1,2,3,4,5,6},
    ytick={0,1,2,3,4,5},
    grid=major,
    grid style={dashed, gray!20},
    thick,
    x=2cm, y=2cm,
    compat=1.11,
    axis lines=left,
    clip=false 
]

    \path (axis cs:1,3) coordinate (A1)
          ++(axis direction cs:2,0)  coordinate (B1)
          ++(axis direction cs:1,-1) coordinate (C1)
          ++(axis direction cs:2,0)  coordinate (D1);

    \draw[thick, myBlue] (A1) -- (B1) node[pos=0, anchor=south west] {firm 1};
    \draw[dotted, myBlue] (B1) -- (C1)
        node[pos=0, sloped, anchor=south west] {$1 \to 2$};
    \draw[thick, myBlue] (C1) -- (D1) 
        node[pos=0, anchor=south west] {firm 2}
        node[pos=1, anchor=west] {worker 1};

    \path (axis cs:1,2) coordinate (A2)
          ++(axis direction cs:1,0)  coordinate (B2)
          ++(axis direction cs:1,-1) coordinate (C2)
          ++(axis direction cs:3,0)  coordinate (D2);

    \draw[thick,myGray] (A2) -- (B2) node[pos=0, anchor=south west] {firm 1};
    \draw[dotted,myGray] (B2) -- (C2)
        node[pos=0, sloped, anchor=south west] {$1 \to 2$};
    \draw[thick,myGray] (C2) -- (D2) 
        node[pos=0, anchor=south west] {firm 2}
        node[pos=1, anchor=west] {worker 2};

    \path (axis cs:1,2.5) coordinate (A3)
          ++(axis direction cs:3,0)  coordinate (B3)
          ++(axis direction cs:1,1) coordinate (C3)
          ++(axis direction cs:1,0)  coordinate (D3);

    \draw[thick,myRed] (A3) -- (B3) node[pos=0, anchor=south west] {firm 2};
    \draw[dotted,myRed] (B3) -- (C3)
        node[pos=0, sloped, anchor=south west] {$2 \to 1$};
    \draw[thick,myRed] (C3) -- (D3) 
        node[pos=0, anchor=south west] {firm 1}
        node[pos=1, anchor=west] {worker 3};

\end{axis}
\end{tikzpicture}
\end{center}
\vspace{-0.5cm}


{\footnotesize
			\textit{Notes:} The figure plots the wage trajectories of three workers with various $\alpha_i$ values ($\alpha_1>\alpha_2$ and $\alpha_1<\alpha_3$) moving through two firms with different $\psi_j$ values ($\psi_2<\psi_1$). Time is on the x axis, and log wages are on the y axis. Colors indicate workers.
		}

   \vspace{0.5cm} 

\end{figure}
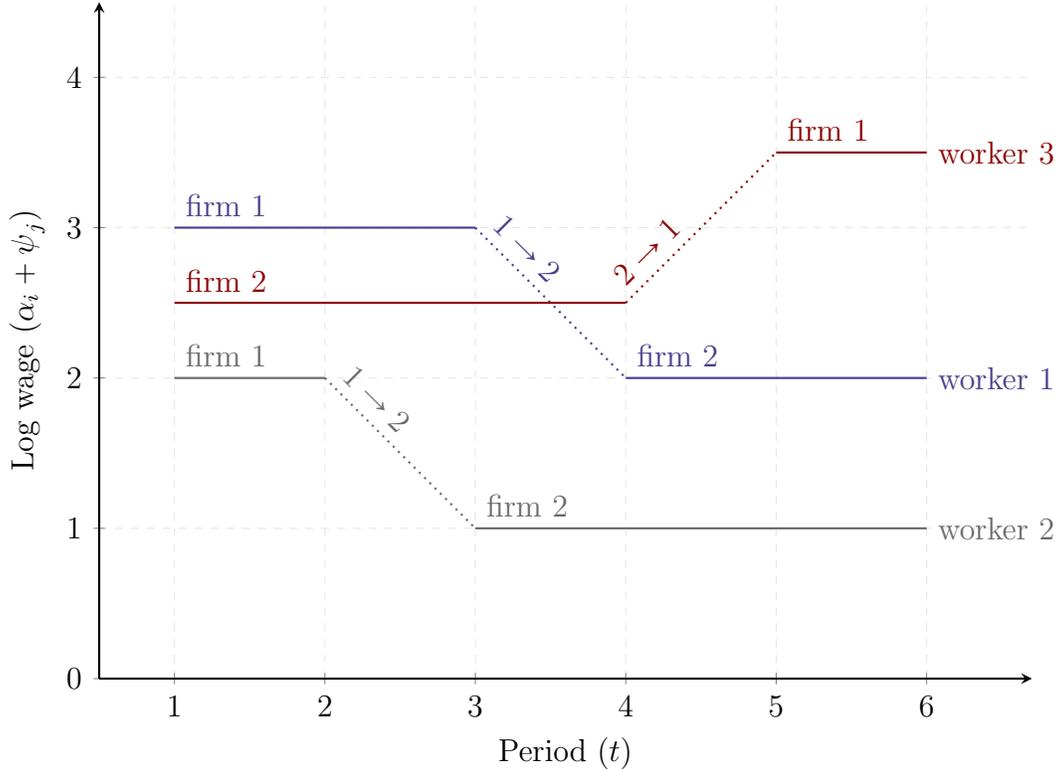

\vskip .3cm \noindent Figure \ref{fig_eventstudy} illustrates the main implications of the model by plotting the log wage over time, for three workers. Initially, in period 1, workers 1 and 2 are employed in firm 1, and earn different wages {due to $\alpha_1$ being greater than $\alpha_2$. When worker 2 moves to firm 2, she experiences a wage reduction since $\psi_2$ is lower than $\psi_1$} -- that is, firm 2 pays similar workers less than firm 1. When, in a later period, worker 1 also moves to firm 2, she experiences exactly the same wage drop in percentage terms as worker 2. Moreover, the wage gap between workers 1 and 2 in firm 2 is the same as in firm 1. Next, worker 3, who was initially employed in the lower-paying firm 2, moves to the better-paying firm 1. The wage gain she experiences exactly mirrors the losses experienced by workers 1 and 2 in their respective job moves.     

%

\vskip .3cm \noindent At its core, the AKM methodology leverages information from workers moving between firms, extracted from linked employer-employee data, to separately identify {the firm components $\psi_j$ and the worker components $\alpha_i$}. Given estimates of these parameters, which are commonly referred to as worker and firm ``effects'', researchers can assess how much of the cross-sectional wage variation and its evolution over time can be explained by worker heterogeneity, firm heterogeneity, and the sorting between workers and firms. 


\vskip .3cm \noindent  {The AKM approach has been influential in labor economics. In this article, we review key empirical findings in the literature while exploring the subtleties of the model and its estimator. We will highlight best practices for overcoming established estimation hurdles and discuss the methodological challenges that still remain, particularly regarding the static and additive nature of the model. }

\vskip .3cm \noindent Throughout the article, we will emphasize that the apparent simplicity of the AKM approach is deceptive. The model does not restrict how job mobility relates to worker and firm heterogeneity, and it can speak to economic models of job choice and wage determination. This explains the growing popularity of AKM in applied work. {Naturally, this sophistication introduces complexity;} the large number of parameters and the network structure of the worker--firm data create important challenges for the estimation of the model and the development of extensions that relax some of its most restrictive assumptions. 

\vskip .3cm \noindent {The AKM methodology was originally developed in the context of wages, workers, and firms, and we will use this setting to illustrate the approach throughout our presentation. {Because many linked datasets share a similar structure, researchers have successfully adapted the model to a wide diversity of questions}. We will mention applications to health economics and corporate finance, but note that the AKM approach has been applied to a diversity of questions in a variety of fields.}
  

\vskip .3cm \noindent Our discussion is intended to provide a concise introduction to the AKM methodology and a users' guide for implementation. For more details, we refer readers to the specialized surveys by \citet{abowd2008econometric}, \citet{card2018firms}, \citet{bonhomme2020econometric}, and \citet{kline2024firm}. 

\vskip .3cm \noindent Lastly, to facilitate the understanding and adoption of the methods, a notebook available \href{https://github.com/tlamadon/abc-of-akm}{here} accompanies the article.

\section*{A Conceptual Framework for AKM}

\vskip .3cm \noindent A useful way to think about the AKM framework introduced by \citet{abowd1999high} is as a simple model of wage determination. Workers are characterized by a fixed level of productivity, $\alpha_i$, and firms are characterized by a pay policy, $\psi_j$, representing how much the firm pays per unit of effective labor. Wages reflect the combination of these two components. The structure is intentionally parsimonious: 
one part of pay is tied to who the worker is, the other to where the worker is employed.

\vskip .3cm \noindent  The worker component, $\alpha_i$, can be interpreted broadly. It may represent something concrete, such as task completion speed, or something more abstract, such as overall labor market ability. 
If two workers have productivities $\alpha_1$ and $\alpha_2$, the difference $\alpha_1 - \alpha_2$ 
reflects the percentage gap in the value they would generate over a fixed period. In this sense, 
$\alpha_i$ summarizes all persistent, worker-specific determinants of earnings.

\vskip .3cm \noindent The firm component, $\psi_j$, captures systematic differences in pay across employers. 
{A high-$\psi_j$ firm may share rents with workers through bargaining or profit-sharing, or compensate workers for less desirable 
working conditions. 
In turn, a low-$\psi_j$ firm may be exercising monopsony power.} The framework does not commit to a particular mechanism; instead, it treats 
$\psi_j$ as a reduced-form measure of firm-level wage-setting. Differences in $\psi_j$ therefore 
correspond to wage premia that apply to any worker employed at that firm.

\vskip .3cm \noindent In logarithms, wages are additively separable as $\alpha_i + \psi_j$. This additive structure yields transparent comparative statics. Holding the firm fixed, a worker with higher $\alpha_i$ earns 
proportionally more than a worker with lower $\alpha_i$. Holding the worker fixed, differences 
in pay across firms equal the difference in their firm effects $\psi_j$. The model thus provides a 
simple accounting framework for wage dispersion across both workers and firms.

\vskip .3cm \noindent To complete the picture, one must describe how workers and firms match together. Workers may survey multiple firms and choose the one offering the highest wage. Alternatively, they may 
encounter firms sequentially and decide whether to move based on wages and nonpecuniary 
considerations. The AKM framework remains largely agnostic about these mechanisms. It allows 
for sorting -- for example, high-productivity workers may be disproportionately employed at high-paying firms -- and for mobility over time, provided that worker and firm compnents are stable and enter wages additively.

\vskip .3cm \noindent Observed wages, of course, do not perfectly equal $\alpha_i + \psi_j$. An empirical specification therefore includes a residual term. In its simplest form, this residual is treated as unrelated 
to the underlying matching process -- a {luck of the draw} rather than an object of economic 
interest. More elaborate interpretations are possible, but the baseline specification abstracts 
from them.

\vskip .3cm \noindent The model is estimated using linked employer--employee data. Each 
observation corresponds to a worker $i$ in period $t$. {We will denote as $j_{it}$ the firm that employs worker $i$ in period $t$, and the observed 
compensation as $Y_{it}$ (in logs).} Because workers move across firms over time, these data allow 
researchers to disentangle persistent worker effects from firm-specific wage premia, and thereby 
decompose overall wage dispersion into its constituent components.

\vskip .3cm \noindent The setting we will use as a running illustrative example throughout the paper concerns the role of firms in shaping wage inequality, which is a central question addressed in the AKM literature. A key tool is a decomposition of wage dispersion into components that reflect dispersion in workers' $\alpha_i$ (\emph{worker heterogeneity}), firms' $\psi_j$ (\emph{firm heterogeneity}), and whether high-$\alpha_i$ workers tend to work in high-$\psi_j$ firms (\emph{sorting}). In the cross-section, these decompositions are informative about the share of variance explained by firms and workers in the overall wage distribution. When applied to various countries, or to various sub-periods in the same country, they shed light on how the role of firm heterogeneity, worker heterogeneity, and sorting, vary across contexts and over time.

\section*{The Econometric Model}

\vskip .3cm \noindent In the AKM model introduced by \citet{abowd1999high}, workers $i$ are characterized by their components, or ``effects'', $\alpha_i$, and firms $j$ are characterized by their components, or ``effects'', $\psi_j$. The model postulates the following expression for the wage (in logs) that worker $i$ would earn at time $t$ if she was employed in firm $j$:
\begin{equation*}Y_{it}(j)=X_{it}\beta+\alpha_i+\psi_{j}+U_{it}.\end{equation*}

\vskip .3cm \noindent The wage depends on the sum of the worker and firm components, $\alpha_i+\psi_j$. As a result, the difference in the wages that a worker would earn in firms $1$ and $2$ is precisely equal to the difference in their components $\psi_1-\psi_2$. This wage differential between the two firms is the same for all workers, and it is constant over time. Hence, in Figure \ref{fig_eventstudy}, workers 1 and 2 experienced exactly the same wage drop when moving, despite the fact that they moved in different periods.   

\vskip .3cm \noindent The actual wage in AKM is not {strictly} equal to the sum of worker and firm components, for two reasons. First, wages also depend on some idiosyncratic shocks $U_{it}$. These shocks account for variation over time in a worker's wage even if she remains in the same firm. Second, wages depend in addition on some covariates $X_{it}$ that are observed in the data -- such as the labor market experience of the worker -- with their associated coefficient $\beta$. 

\vskip .3cm \noindent The quantity $Y_{it}(j)$ is the wage that would result from the worker being exogenously assigned to firm $j$. Hence, for any given worker in the economy, in any given period, the wage equation specifies \emph{all} the wages that the worker could have earned in \emph{all} possible firms. However, the data is not directly informative about those potential wages. The observed wage in the sample, say $Y_{it}$, corresponds to the firm that worker $i$ is employed at in period $t$. {Since we have denoted this firm as $j_{it}$, the wage that is recorded in the data is thus $Y_{it}=Y_{it}(j_{it})$.}

\vskip .3cm \noindent It is important to note that there are typically many thousands of firms in the sample, and the data only provide information about the wage at one firm in a given period. Hence, recovering potential wages in all firms requires solving a formidable extrapolation problem. Remarkably, the structure of the AKM model gives researchers the ability to predict the wages of a worker in all (connected) firms in the economy. In practice, this is achieved through the use of the AKM estimator that we will describe in the next section.

\vskip .3cm \noindent Under the AKM model, the average wage in the firm {reflects a combination of firm and worker effects}. Abstracting from covariates and idiosyncratic shocks for simplicity, the average wage in firm $j$ is the sum of the firm component $\psi_j$, and the average of the worker components $\alpha_i$ for the workers employed in the firm. A potential explanation for a high mean wage in the firm is that it has a high $\psi_j$, through which it consistently pays its workers a high wage. However, an alternative explanation for a high wage is that workers in the firm have high $\alpha_i$, meaning that the firm employs workers who would earn good wages irrespective of where they work. {Hence, average wages in the firm are not sufficient to disentangle firm and worker components.}

\vskip .3cm \noindent {The key feature of the model that allows researchers to separately recover worker and firm effects is job mobility, under the assumption that the latter is \emph{exogenous}.} As we indicated in the conceptual framework, the AKM model does not restrict how worker--firm matches, here denoted as $j_{it}$, are related to the worker and firm components, $\alpha_i$ and $\psi_j$. At the same time, the model does impose that the shock $U_{it}$ be uncorrelated with the set of all worker--firm matches in the data. This assumption, which is referred to as \emph{exogenous mobility} in the literature, is central to the methodology. Under exogenous mobility, the difference in average wages of workers who switch between firms is directly informative about the firms' components. While some evidence has been offered in support of exogenous mobility, the assumption has been debated and remains controversial. We will mention some limitations of the model's assumptions, as well as extensions of the model, in the last section of the article.

\vskip .3cm \noindent Taking stock, in the AKM setup, the worker's wage $Y_{it}$ at match $j_{it}$ is a linear function of the worker component, the firm component, and the covariates. Moreover, under the exogenous mobility assumption, the error term $U_{it}$ is unrelated to the worker--firm matches. The Ordinary Least Squares (OLS) estimator is the most popular estimator in such settings, and it is the estimator that the AKM approach relies on. We present the estimator in the next section. 

\vskip .3cm \noindent Returning to our leading example, \citet{card2013workplace} study how firm and worker heterogeneity shape the structure and evolution of wages in Germany. In their sample, 16 million workers are followed during the 2002--2009 period. The AKM model provides predictions of the wages those workers would earn in all 1,500,000 firms (establishments) in the sample. {For this reason, the authors argue that it is important to offer evidence in support of the model's assumptions, and they report a number of diagnostic checks to this end. In the last section of the article, we will mention several diagnostics and extensions of the AKM model.}


\section*{The AKM Estimator}

\vskip .3cm \noindent In this section we describe the AKM estimator introduced by \citet{abowd1999high}. The starting point is that the researcher has access to linked employer-employee data. These data contain a panel component, since they follow workers over time, as well as a firm component, since they keep track of the firm identifiers. Linked employer-employee data sets are now available in many countries.


\vskip .3cm \noindent Estimating a linear regression in typical applications of AKM requires handling the presence of a large number of parameters. In many linked data sets, there are hundreds of thousands or millions of workers and firms, hence the same number of worker- and firm-specific parameters to estimate. While initially challenging, estimation algorithms are now well understood, and efficient computational routines have been developed. A state-of-the-art computer package, which also includes the improvements to the original AKM approach that we will discuss in the last two sections of this article, is \emph{pytwoway}, available \href{https://tlamadon.github.io/pytwoway/}{here}. 

\vskip .3cm \noindent A specific feature of the AKM setting is that the variables on the right-hand side of the regression are not linearly independent. This matters, since a ``naive'' regression will simply not produce a number. Dependence between right-hand-side variables arises mechanically since, in the wage equation, only the sum of worker and firm components can be identified but their separate levels are not. Hence, one cannot hope to recover the average $\alpha_i$ separately from the average $\psi_j$. This issue is easy to address by normalizing one of the coefficients, such as the average firm component in the sample, to zero.  

\vskip .3cm \noindent Another reason why right-hand-side variables are linearly dependent is related to the \emph{lack of connectivity} of the worker--firm network. Consider the case where the workers in a firm never leave the firm, and the firm does not hire new workers during the entire observation period. In this case, there is no way to know whether the workers have high $\alpha_i$ or the firm has a high $\psi_j$, and consequently the firm's $\psi_j$ is not identified. More generally, identification is only secured within \emph{connected sets} of the worker--firm graph. In such a set, it is possible to link any two firms by tracing out workers' movements. 

\vskip .3cm \noindent {For example, to identify the $\psi_j$ components of firms 1 and 2 one can compare the wages of workers who are employed at both firms at some point in the sample (i.e., $1\leftrightarrow 2$ movers). Alternatively, if no worker moves between 1 and 2, one can consider another firm, say firm 3, and compute the wage difference for $1\leftrightarrow 3$ movers, and subtract off the wage difference for $2\leftrightarrow 3$ movers.}

\vskip .3cm \noindent The leading approach to deal with lack of connectivity is to focus on a single (typically, the largest) connected set of the worker--firm graph, often called ``the'' connected set. \citet{abowd2002computing} propose a simple algorithm to compute {connected sets}, which works as follows. Starting by including a worker in the set, include all the firms where the worker has been employed at some point during the sample period. Then, include in the set all the workers who have been employed at some point in those firms. This process is continued until the size of the set ceases to increase. In many applications, the largest connected set is large and contains most workers and most moderate and large firms, yet it typically leaves out some small firms and their workers.    

\vskip .3cm \noindent When focusing on the connected set, computing the AKM estimator requires solving a linear system. This is a large, yet sparse, linear system, since workers only visit a handful of firms during the period. Reliable computational routines with minimal memory requirements have been developed, notably the iterative ``zigzag'' algorithm of \citet{guimaraes2010simple}.\footnote{An alternative estimation approach is to estimate the AKM model in first differences, exploiting that $Y_{it}-Y_{i,t-1}$ no longer depends on $\alpha_i$. One can then estimate the $\psi_j$ parameters using least squares, and recover the $\alpha_i$ parameters after the fact.}  

\vskip .3cm \noindent The zigzag algorithm works as follows. Instead of trying to jointly recover the parameters $\alpha_i$, $\psi_j$, and $\beta$, the algorithm recovers them one at a time in an iterative fashion. Indeed, given $\psi_j$ and $\beta$, $\alpha_i$ can be estimated as a simple worker-specific average. In turn, given $\alpha_i$ and $\beta$, $\psi_j$ can be estimated as a firm-specific average. Lastly, given $\alpha_i$ and $\psi_j$, $\beta$ can be efficiently estimated by standard linear regression (since $\beta$ is a ``usual'', low-dimensional parameter). Starting from some initial values, iterating between these three sets of parameters until convergence provides a fast and reliable estimation routine.


\vskip .3cm \noindent The most popular use of the AKM methodology is as a way to quantify variance components, {using} the following decomposition of the total variance of log wage residuals:
\begin{equation*}
	\underset{\text{total log wage variance}}{\limfunc{Var}(Y_{it}-X_{it}\beta)}=	\underset{\text{worker variance}}{\limfunc{Var}(\alpha_i)}+\underset{\text{firm variance}}{\limfunc{Var}(\psi_{j_{it}})}+2\underset{\text{covariance}}{\limfunc{Cov}(\alpha_i,\psi_{j_{it}})}+\underset{\text{idiosyncratic variance}}{\limfunc{Var}(\varepsilon_{it})}.
\end{equation*}

\vskip .3cm \noindent In this decomposition, the worker and firm components quantify how much of the dispersion in log wages can be attributed to dispersion in worker and firm effects, respectively. For example, in an economy where firms pay similar workers identically, $\psi_j$ is constant among all firms and its variance is equal to zero. In such an economy, wage differences between firms solely reflect differences in the types of workers they employ.

\vskip .3cm \noindent An important term in the decomposition is the covariance component. Note that the firm component $\psi_{j}$ is evaluated at the actual worker--firm match $j_{it}$. A positive covariance thus indicates that high-$\alpha_i$ workers tend to work in high-$\psi_j$ firms, while a negative covariance indicates the opposite. This covariance, and the associated correlation coefficient, are commonly interpreted as measuring the contribution of worker--firm \emph{sorting} to the overall dispersion in log wages. Consistently with this interpretation, if workers moved randomly between firms and employment patterns were independent of worker and firm heterogeneity, then the covariance component would be equal to zero.         

\vskip .3cm \noindent Decompositions of this form are commonly reported in empirical work. A direct extension is to decompose the total variance of log wages that includes covariate effects, by adding variance and covariance terms quantifying the contribution of covariates. Another common approach is to decompose the between-firm variance in addition to the total variance.

\vskip .3cm \noindent Returning to our example on the role of firms in shaping wage inequality, an important goal in the literature has been to assess the quantitative magnitude of worker variance, firm variance, and covariance in explaining wage dispersion. For example, \citet{card2013workplace} and \citet{song2019firming} report estimates of variance components in the cross-section, and study how they evolve over time and help explain the evolution of wage inequality in Germany and the US, respectively. We will mention some of their main empirical findings in the next section.

\vskip .3cm \noindent A common approach to estimation is to compute the variances and covariances of the estimated worker and firm effects $\alpha_i$ and $\psi_j$. For example, the estimator of the worker variance proposed by \citet{abowd1999high} is simply the variance of the estimated worker parameters $\alpha_i$. However, in many empirical applications of AKM, this approach leads to an estimator with a potentially large \emph{bias}. In the penultimate section of the article we will mention alternative estimators of variances and covariances that correct for the bias and lead to more reliable estimates.

\vskip .3cm \noindent In addition to variance decompositions, the AKM method is commonly used for a variety of other purposes in the literature. In some settings, researchers are interested in the coefficient $\beta$ of some covariate. In this case, the presence of worker and firm components on the right-hand-side of the regression {captures} potential confounding factors. As an example, \citet{lavetti2016estimating} control for worker and firm effects when estimating compensating wage differentials for occupational fatality risk, thus accounting for the possibility that fatality risk may differ across workers and firms.

\vskip .3cm \noindent In other settings, researchers are interested in averages of firm components $\psi_j$ for various groups of firms. As an example, \citet{setzler2021effects} compare averages of $\psi_j$ parameters for multinational and national firms. This methodology accounts for the fact that firms are heterogeneous within groups, while controlling for worker heterogeneity as well. Another example is the analysis and decomposition of inter-industry wage differentials pioneered by \citet{abowd1999high}, and studied recently in \citet{card2024industry}.


\vskip .3cm \noindent More broadly, AKM estimates of worker and firm components are also commonly used on the left-hand-side, or on the right-hand-side, of regressions. In addition, researchers often correlate worker and firm effects with other characteristics of workers and firms in order to relate them to interpretable dimensions that are measured in the data. Recovering AKM estimates of worker and firm components, to later use them in a subsequent part of the analysis, has become ubiquitous in applied labor economics. 

\section*{Some Applications of AKM}

\vskip .3cm \noindent Since its introduction in \citet{abowd1999high}, the AKM methodology has been widely used to produce estimates of worker and firm effects, and to shed light on features of labor markets. Here we review several findings reported in the literature, and then mention some applications of AKM to other fields within economics.


\vskip .3cm \noindent The starting point, and the motivation, for the analysis is the large share of wage dispersion that occurs \emph{between firms}. In the US and continental Europe, for example, the between-firm share of variance in log wages typically accounts for between 30\% and 50\% of the total variance. This indicates that wages differ sharply between firms, not only within. However, through the lens of the AKM model, these wage differences may reflect several distinct mechanisms.

\vskip .3cm \noindent Consider an extreme case where firms pay workers equally, i.e., the $\psi_j$ components are identical in all firms in the economy. It is possible to reconcile this hypothetical scenario with large wage differences between firms if they employ different types of workers. Although firms pay \emph{similar workers} the same since $\psi_j$ is constant, they pay \emph{the workers they employ} differently since some firms employ high-$\alpha_i$ workers while other firms employ low-$\alpha_i$ workers. At the other extreme, substantial wage differences between firms are also consistent with workers being identical in the sense that their $\alpha_i$ components are the same, yet high-$\psi_j$ and low-$\psi_j$ firms pay workers differently.

\vskip .3cm \noindent The AKM model, and the use of linked worker--firm data, allow researchers to identify these distinct mechanisms and shed light on the sources of the wage differences between and within firms. The literature has studied both the sources of cross-sectional wage dispersion and the factors explaining the evolution of wage inequality over time.

\vskip .3cm \noindent In the cross-section, all three shares of variance -- that is, worker variance, firm variance, and covariance -- have been found to explain sizable shares of the overall wage dispersion, although there is substantial variation in estimated relative shares across settings and periods. Across a set of empirical studies, \citet{bonhomme2023much} report that the interquartile range of the percentage of variance explained by firm components is 15\% to 25\%, while the corresponding range for the percentage of variance explained by the covariance term includes estimates larger than 15\% as well as some negative estimates.

\begin{figure}[h!]
	\caption{Decomposition of the variance of log annual earnings in the US}
	\label{fig:vardec}
	\vspace{-15pt}
		
	\centering
    \begin{tikzpicture}[scale=0.8, yscale=0.7]
		
	\definecolor{bandblue}{HTML}{5C7B97}
	\definecolor{bandyellow}{RGB}{158, 202, 225} 
	\definecolor{workerD}{RGB}{107, 174, 214}    
	\definecolor{bandpink}{RGB}{252, 146, 114}   
	\definecolor{firmD}{RGB}{251, 106, 74}       
	\definecolor{bandgreen}{RGB}{161, 217, 155}  
	\definecolor{sortD}{RGB}{116, 196, 118}      
	\definecolor{bandbrown}{RGB}{189, 189, 189}  
	\definecolor{residD}{RGB}{150, 150, 150}     
    
    \pgfkeys{/pgf/number format/.cd, fixed, fixed zerofill, precision=3}
		
	\useasboundingbox (1.5, -4) rectangle (11, 8);

    
    \pgfmathsetmacro{\vTot}{0.801}    
    
    \pgfmathsetmacro{\vBet}{0.309}    
    \pgfmathsetmacro{\vWith}{0.492}   
    
    \pgfmathsetmacro{\vFirm}{0.081}   
    \pgfmathsetmacro{\vCov}{0.108}    
    \pgfmathsetmacro{\vWorker}{0.476} 
    \pgfmathsetmacro{\vResid}{0.136}  

    \pgfmathsetmacro{\S}{10}   
    \pgfmathsetmacro{\gm}{1.0} 
    \pgfmathsetmacro{\gr}{1.2} 

    \def\xLeft{0} \def\xMid{6} \def\xRight{13}
    \def\cpA{3} \def\cpB{9.5}

    \pgfmathsetmacro{\yLTop}{\S * \vTot / 2}
    \pgfmathsetmacro{\yLBot}{-\S * \vTot / 2}
    \pgfmathsetmacro{\yLSplit}{\yLTop - \S * \vBet}

    \pgfmathsetmacro{\yMTop}{\S * \vTot / 2 + \gm / 2}
    \pgfmathsetmacro{\yMBetBot}{\yMTop - \S * \vBet}
    \pgfmathsetmacro{\yMWithTop}{\yMBetBot - \gm}
    \pgfmathsetmacro{\yMBot}{\yMWithTop - \S * \vWith}

    \pgfmathsetmacro{\yMBetSplitA}{\yMTop - \S * \vFirm}
    \pgfmathsetmacro{\yMBetSplitB}{\yMBetSplitA - \S * \vCov}
    \pgfmathsetmacro{\yMWithSplit}{\yMWithTop - \S * (\vWith - \vResid)}

    \pgfmathsetmacro{\yRTop}{\S * \vTot / 2 + 1.5 * \gr}
    \pgfmathsetmacro{\yRFirmBot}{\yRTop - \S * \vFirm}
    \pgfmathsetmacro{\yRCovTop}{\yRFirmBot - \gr}
    \pgfmathsetmacro{\yRCovBot}{\yRCovTop - \S * \vCov}
    \pgfmathsetmacro{\yRWorkTop}{\yRCovBot - \gr}
    \pgfmathsetmacro{\yRWorkSplit}{\yRWorkTop - \S * (\vBet - \vFirm - \vCov)}
    \pgfmathsetmacro{\yRWorkBot}{\yRWorkSplit - \S * (\vWith - \vResid)}
    \pgfmathsetmacro{\yRResTop}{\yRWorkBot - \gr}
    \pgfmathsetmacro{\yRResBot}{\yRResTop - \S * \vResid}

    
    \fill[bandblue] 
        (\xLeft, \yLTop) .. controls (\cpA, \yLTop) and (\xMid-\cpA, \yMTop) .. (\xMid, \yMTop) --
        (\xMid, \yMBetBot) .. controls (\xMid-\cpA, \yMBetBot) and (\cpA, \yLSplit) .. (\xLeft, \yLSplit) -- cycle;
        
    \fill[bandblue] 
        (\xLeft, \yLSplit) .. controls (\cpA, \yLSplit) and (\xMid-\cpA, \yMWithTop) .. (\xMid, \yMWithTop) --
        (\xMid, \yMBot) .. controls (\xMid-\cpA, \yMBot) and (\cpA, \yLBot) .. (\xLeft, \yLBot) -- cycle;

    \fill[bandpink] 
        (\xMid, \yMTop) .. controls (\cpB, \yMTop) and (\cpB, \yRTop) .. (\xRight, \yRTop) --
        (\xRight, \yRFirmBot) .. controls (\cpB, \yRFirmBot) and (\cpB, \yMBetSplitA) .. (\xMid, \yMBetSplitA) -- cycle;

    \fill[bandgreen] 
        (\xMid, \yMBetSplitA) .. controls (\cpB, \yMBetSplitA) and (\cpB, \yRCovTop) .. (\xRight, \yRCovTop) --
        (\xRight, \yRCovBot) .. controls (\cpB, \yRCovBot) and (\cpB, \yMBetSplitB) .. (\xMid, \yMBetSplitB) -- cycle;

    \fill[bandyellow] 
        (\xMid, \yMBetSplitB) .. controls (\cpB, \yMBetSplitB) and (\cpB, \yRWorkTop) .. (\xRight, \yRWorkTop) --
        (\xRight, \yRWorkSplit) .. controls (\cpB, \yRWorkSplit) and (\cpB, \yMBetBot) .. (\xMid, \yMBetBot) -- cycle;

    \fill[bandyellow] 
        (\xMid, \yMWithTop) .. controls (\cpB, \yMWithTop) and (\cpB, \yRWorkSplit) .. (\xRight, \yRWorkSplit) --
        (\xRight, \yRWorkBot) .. controls (\cpB, \yRWorkBot) and (\cpB, \yMWithSplit) .. (\xMid, \yMWithSplit) -- cycle;

    \fill[bandbrown] 
        (\xMid, \yMWithSplit) .. controls (\cpB, \yMWithSplit) and (\cpB, \yRResTop) .. (\xRight, \yRResTop) --
        (\xRight, \yRResBot) .. controls (\cpB, \yRResBot) and (\cpB, \yMBot) .. (\xMid, \yMBot) -- cycle;

    \tikzset{
        midlabel/.style={
            fill=white, fill opacity=0.9, text opacity=1, 
            rounded corners=3pt, align=center, inner sep=5pt
        }
    }
    
    \pgfmathsetmacro{\yMLabelBet}{(\yMTop + \yMBetBot)/2}
    \pgfmathsetmacro{\yMLabelWith}{(\yMWithTop + \yMBot)/2}
    \pgfmathsetmacro{\yRLabelFirm}{(\yRTop + \yRFirmBot)/2}
    \pgfmathsetmacro{\yRLabelCov}{(\yRCovTop + \yRCovBot)/2}
    \pgfmathsetmacro{\yRLabelWork}{(\yRWorkTop + \yRWorkBot)/2}
    \pgfmathsetmacro{\yRLabelRes}{(\yRResTop + \yRResBot)/2}

    \node[anchor=east, align=right] at (-0.2, 0) {Total Variance\\[-0.2em] \large \pgfmathprintnumber{\vTot}};
    
    \node[midlabel] at (\xMid, \yMLabelBet) {Between-Firm Variance\\[-0.2em] \large \pgfmathprintnumber{\vBet}};
    \node[midlabel] at (\xMid, \yMLabelWith) {Within-Firm Variance\\[-0.2em] \large \pgfmathprintnumber{\vWith}};
    
    \node[anchor=west, align=left] at (\xRight + 0.2, \yRLabelFirm) {Firm Variance\\[-0.2em] \large \pgfmathprintnumber{\vFirm}};
    \node[anchor=west, align=left] at (\xRight + 0.2, \yRLabelCov) {Covariance\\[-0.2em] \large \pgfmathprintnumber{\vCov}};
    \node[anchor=west, align=left] at (\xRight + 0.2, \yRLabelWork) {Worker Variance\\[-0.2em] \large \pgfmathprintnumber{\vWorker}};
    \node[anchor=west, align=left] at (\xRight + 0.2, \yRLabelRes) {Residual\\[-0.2em] \large \pgfmathprintnumber{\vResid}};

\end{tikzpicture}
	
	\begin{quote}
		{\footnotesize
			\textit{Notes:} The figure is constructed from Tables III and IV in \citet{song2019firming}. The numbers correspond to the $2007-2013$ period, and are net of observed covariates. Worker variance is indicated in blue, firm variance in red, covariance in green, and residual variance in gray. 
		}
	\end{quote}
\end{figure}
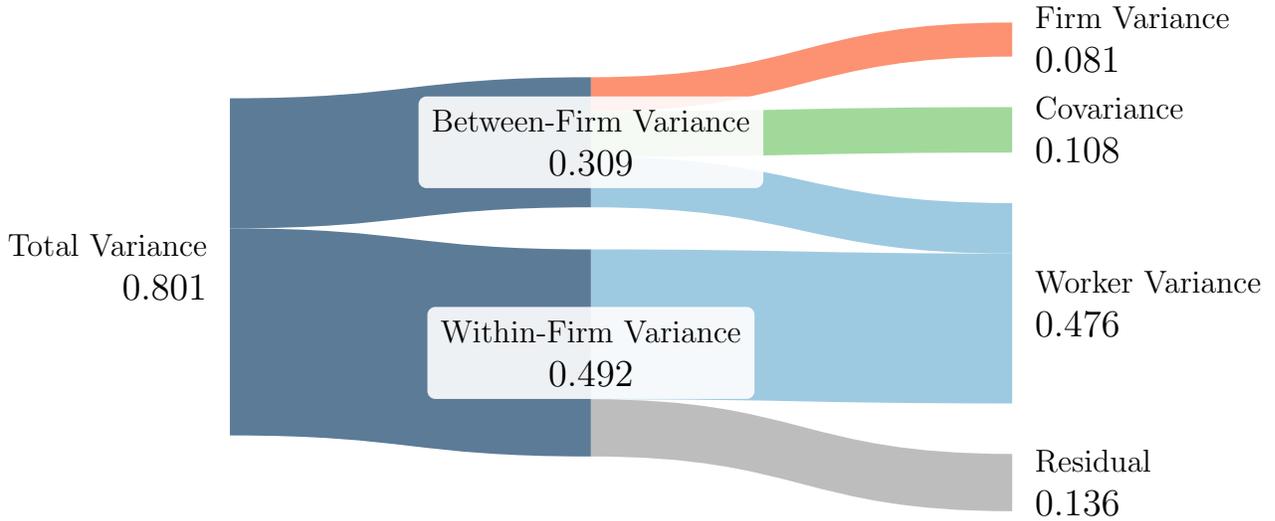

\vskip .3cm \noindent As an example, in Figure \ref{fig:vardec} we report estimates from \citet{song2019firming}, based on the US for the 2007--2013 period. The figure shows that the between-firm variance amounts to 39\% of the total variance of log annual earnings. The AKM decomposition gives a sharper conclusion: 10\% of the variation is explained by firms, 59\% by workers, and 13\% by the covariance, the remainder being explained by idiosyncratic shocks. Note that, as Figure \ref{fig:vardec} illustrates, while the firm variance and covariance arise solely from between-firm dispersion, the worker variance reflects a combination of within- and between-firm dispersion.

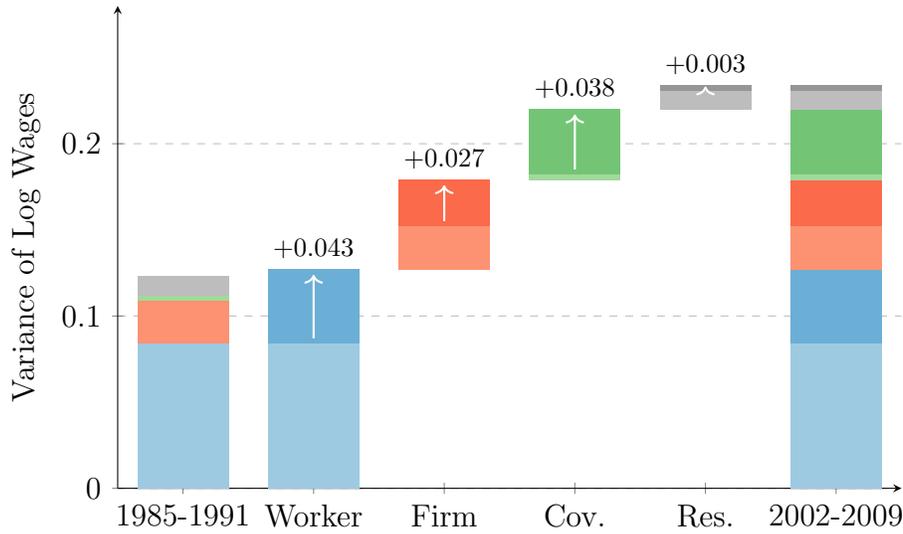
\begin{figure}[h!]
	\caption{Change in the decomposition of the variance of log daily wages over time in Germany}
	\label{fig:vardec-chk}
		
	\centering
    \vspace{0.5cm}
    \begin{tikzpicture}
	
	\definecolor{workerL}{RGB}{158, 202, 225} 
	\definecolor{workerD}{RGB}{107, 174, 214} 
	\definecolor{firmL}{RGB}{252, 146, 114}   
	\definecolor{firmD}{RGB}{251, 106, 74}    
	\definecolor{sortL}{RGB}{161, 217, 155}   
	\definecolor{sortD}{RGB}{116, 196, 118}   
	\definecolor{residL}{RGB}{189, 189, 189}  
	\definecolor{residD}{RGB}{150, 150, 150}  

	\useasboundingbox (-1.5, 1) rectangle (11, 6);
	
	\begin{axis}[
		width=12cm, height=8cm,
		ymin=0, ymax=0.28,
		xmin=0.5, xmax=6.5,
		xtick={1,2,3,4,5,6},
		xticklabels={$1985\text{-}1991$, Worker, Firm, Cov., Res., $2002\text{-}2009$},
        ytick={0, 0.1, 0.2, 0.3},
		ylabel={Variance of Log Wages},
		axis x line=bottom,
		axis y line=left,
		enlarge x limits=false,
		ymajorgrids=true,
		grid style=dashed,
		title style={font=\bfseries, yshift=2ex}
		]
		
		\def\hw{0.35} 
		
		\fill[workerL] (1-\hw, 0)     rectangle (1+\hw, 0.084);
		\fill[firmL]   (1-\hw, 0.084) rectangle (1+\hw, 0.109);
		\fill[sortL]   (1-\hw, 0.109) rectangle (1+\hw, 0.112);
		\fill[residL]  (1-\hw, 0.112) rectangle (1+\hw, 0.123);
		
		\fill[workerL] (2-\hw, 0)     rectangle (2+\hw, 0.084); 
		\fill[workerD] (2-\hw, 0.084) rectangle (2+\hw, 0.127); 
		\node[font=\footnotesize, above] at (2, 0.127) {+0.043};
		
		\draw[->, thick, white, shorten >=2pt, shorten <=2pt] (2, 0.084) -- (2, 0.127);
		
		\fill[firmL] (3-\hw, 0.127) rectangle (3+\hw, 0.152); 
		\fill[firmD] (3-\hw, 0.152) rectangle (3+\hw, 0.179); 
		\node[font=\footnotesize, above] at (3, 0.179) {+0.027};
		
		\draw[->, thick, white, shorten >=2pt, shorten <=2pt] (3, 0.152) -- (3, 0.179);
		
		\fill[sortL] (4-\hw, 0.179) rectangle (4+\hw, 0.182);  
		\fill[sortD] (4-\hw, 0.182) rectangle (4+\hw, 0.220); 
		\node[font=\footnotesize, above] at (4, 0.220) {+0.038};
		
		\draw[->, thick, white, shorten >=2pt, shorten <=2pt] (4, 0.182) -- (4, 0.220);
		
		\fill[residL] (5-\hw, 0.220) rectangle (5+\hw, 0.231); 
		\fill[residD] (5-\hw, 0.231) rectangle (5+\hw, 0.234);  
		\node[font=\footnotesize, above] at (5, 0.234) {+0.003};
		
		\draw[->, thick, white, shorten >=0.5pt, shorten <=0.5pt] (5, 0.231) -- (5, 0.234);
		
		\fill[workerL] (6-\hw, 0)     rectangle (6+\hw, 0.084);
		\fill[workerD] (6-\hw, 0.084) rectangle (6+\hw, 0.127);
		\fill[firmL]   (6-\hw, 0.127) rectangle (6+\hw, 0.152);
		\fill[firmD]   (6-\hw, 0.152) rectangle (6+\hw, 0.179);
		\fill[sortL]   (6-\hw, 0.179) rectangle (6+\hw, 0.182);
		\fill[sortD]   (6-\hw, 0.182) rectangle (6+\hw, 0.220);
		\fill[residL]  (6-\hw, 0.220) rectangle (6+\hw, 0.231);
		\fill[residD]  (6-\hw, 0.231) rectangle (6+\hw, 0.234);			
	\end{axis}
\end{tikzpicture}

	\begin{quote}
		{\footnotesize
			\textit{Notes:} The figure is constructed from Table IV in \citet{card2013workplace}. The numbers are net of observed covariates. Worker variance is indicated in blue, firm variance in red, covariance in green, and residual variance in gray.
		}
	\end{quote}
\end{figure}

\vskip .3cm \noindent AKM decompositions are also commonly used to interpret changes in wage inequality over time. As an example, in Figure \ref{fig:vardec-chk} we report changes between 1985--1991 and 2002--2009 in German daily wages, as estimated by \citet{card2013workplace}. Out of a 11.1 percentage points variance increase, the authors estimate that the change in worker variance accounted for 39\%, the change in firm variance accounted for 24\%, and the change in covariance accounted for 34\% of the increase, with a small residual part explained by idiosyncratic shocks. \citet{song2019firming} produced a similar analysis of the changes in earnings inequality in the US between 1978 and 2013, also finding a substantial increase in the covariance term.

\vskip .3cm \noindent However, some of the findings based on the AKM methodology have been found to be fragile. An issue that has been extensively studied is the \emph{bias} of variance shares, due to the imprecise estimation of the worker and firm parameters $\alpha_i$ and $\psi_j$. In the next section, we will explain the source of the bias and review some available approach to alleviate the issue. Importantly, using available correction methods can make material differences to the estimates.


\vskip .3cm \noindent While \citet{abowd1999high} proposed their method to study workers and firms, many other empirical settings have a similar structure, and AKM has been used in a variety of fields. {Domains of application} include the economics of education, health economics, international trade, and corporate finance, among others. 

\vskip .3cm \noindent An example in health economics is \citet{finkelstein2016sources}, who study the sources of geographic differences in health care utilization. The existing disparity could reflect place-specific factors, such as doctors’ incentives or the quality of hospitals. Alternatively, it could be due to differences in patients' levels of sickness or preferences. While the former explanation could be used to justify policies changing doctors' incentives, the latter cannot. To identify these two possible mechanisms, the authors exploit patients' migration across regions using the AKM methodology. They find that both patient-specific components and place-specific components contribute {substantially to variation in utilization across areas, and that the covariance between these two components is positive}.

\vskip .3cm \noindent Another example in the context of corporate finance is \citet{amiti2018much}, who develop a methodology to separately recover bank-specific supply shocks and firm-specific demand shocks using lending data. Similarly to the AKM approach, their methodology accounts for the presence of firm and bank components in a linear regression. However, unlike in labor market settings where workers are not employed in multiple firms in a given period, at each point in time a firm may borrow from various banks -- and, of course, banks lend to multiple firms as well. This permits researchers to estimate time-varying firm and bank parameters, interpreted as shocks. {Using those in an analysis of the determinants of firms' investment decisions, the authors find that bank-specific shocks contribute significantly to aggregate investment fluctuations.}

\section*{The Bias of AKM}

\vskip .3cm \noindent Although the AKM methodology pioneered by \citet{abowd1999high} has been highly influential, it suffers from some important limitations. Those can be separated into two categories. The first one concerns the issues with the AKM estimator, and chiefly the problem of bias. {The second category concerns the model's assumptions, and whether the AKM model provides a good description of actual wages.} We will now review these two types of limitations in this section and the next, while mentioning some extensions of the original approach that aim at providing improvements.

\vskip .3cm \noindent To understand the bias issue, it is important to note that estimates of worker and firm components $\alpha_i$ and $\psi_j$ are commonly contaminated by a substantial amount of \emph{noise}. To shed light on the source of the noise, recall that differences between the $\psi_j$ components of firms are identified by the wages of the workers moving between these firms. {Hence}, in order to reliably estimate firm components $\psi_j$, the number of job movers is key. 

\vskip .3cm \noindent Moreover, as we indicated when we described the AKM estimator, the ability to recover firm and worker components hinges on a \emph{connectivity} condition, which requires that any two firms be connected through job movers. When two firms are not connected -- either directly or indirectly -- it is not possible to recover the relative components $\psi_j$ of these two firms. {Likewise, when two firms are weakly connected through only a handful of job movers, the estimates of the firms' components tend to be noisy.}

\vskip .3cm \noindent Due to insufficient job movers in some firms and an insufficient degree of connectivity, AKM estimates of firm and worker components are often noisy. For many quantities of interest such as variance shares, the noise in worker and firm effects creates a \emph{bias} on estimates of those quantities. Bias arises since estimation noise does not average out for quantities such as variances and covariances. {Importantly, bias may still be substantial despite the presence of very large numbers of workers and firms}. Qualitatively, AKM estimates tend to overestimate the contribution of firms to wage dispersion. In turn, the covariance between worker and firm components is also biased, but the bias tends to be downward so AKM tends to underestimate the contribution of the covariance term. 


\vskip .3cm \noindent {Nevertheless, concerns with bias} do not apply equally to all quantities of interest. For example, if one is interested in the coefficient $\beta$ of a covariate, or in comparing the average firm components in two groups of firms, then the estimation noise is likely to average out. However, it is difficult to know how particular dimensions of the data, such as the number of movers per firm or the degree of connectivity of the network, affect the noise in the AKM estimates and the resulting bias on the quantity of interest (\citealp{jochmans2019fixed}).

\begin{figure}[h!]
	\caption{Sub-sampling experiment (German data)}
	\label{fig:SW-attrition}
	\begin{center}
		\includegraphics[width=.6\linewidth]{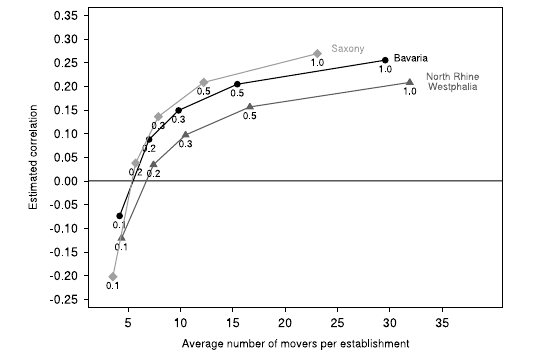}  
\end{center}
	
	\begin{quote}
		{\footnotesize
			\textit{Notes:} The figure is reproduced from \citet{andrews2012high}. The average number of job movers per establishment is shown in the x axis, while the estimate of the correlation between worker and firm effects is shown on the y axis. The three curves correspond to three German regions. 
		}
	\end{quote}
\end{figure}

\vskip .3cm \noindent As a way to assess the importance of bias for the question at hand, we recommend implementing the \emph{sub-sampling plot} proposed by \citet{andrews2012high}. To describe the method, start with a sub-sample of large and well-connected firms. Then, remove a random subset of job movers within each firm. As the share of movers removed increases, the extent of the bias on the quantity of interest is likely to increase since firms in the sub-sample have fewer movers and tend to be less well-connected. The graph simply plots estimates of the quantity of interest on the y axis against the share of job movers removed on the x axis. 

\vskip .3cm \noindent We show an example of a sub-sampling plot in Figure \ref{fig:SW-attrition}, taken from \citet{andrews2012high}, using German data. The authors first select a random sub-sample of $10\%$ workers, indicated as $0.1$ in the figure. They then add back the workers to achieve 20\%, 50\%, and 100\% of the original sample, respectively  (in the largest connected set). Given a share of workers, they report the AKM estimate of the correlation between worker and firm effects on the y axis, and the average number of movers per establishment on the x axis, separately for three German regions.

\vskip .3cm \noindent The figure shows that, in smaller, less connected samples, the correlation is low or negative, while in larger, more connected samples the correlation is positive and quite large. The correlation coefficient is commonly interpreted as reflecting the sign and strength of sorting in the economy, and the figure provides clear evidence of bias that is consistent with AKM-based measures of covariances and correlations being biased downward. Similar findings have been reported for other countries and other quantities of interest, such as the variance of firm effects and the covariance component, see \citet{bonhomme2023much}.

\vskip .3cm \noindent In addition to reporting a sub-sampling plot, we recommend implementing \emph{bias correction methods}, and applying them to the quantity of interest. Available methods to correct the bias can be divided into two categories: fixed-effects methods and correlated random-effects methods. Reliable implementations of both approaches are available in the \href{https://tlamadon.github.io/pytwoway/}{pytwoway} package that we have already mentioned, and we now review them in turn.

\begin{figure}[h!]
	\caption{Impact of bias correction on variance components\label{fig:literature}}
	\begin{center}
        	
\begin{tikzpicture}
	\useasboundingbox (-0.5, 0.5) rectangle (11, 7);

    \begin{axis}[
        width=13cm,
        height=9cm,
        grid=major,
        xmin=-0.15, xmax=0.20,
        ymin=0, ymax=0.30,
        xtick={-0.15, -0.1, -0.05, 0, 0.05, 0.1, 0.15, 0.2},
        xticklabel={\pgfmathparse{\tick*100}\pgfmathprintnumber[fixed, zerofill, precision=0]{\pgfmathresult}\%},
        ytick={0, 0.05, 0.1, 0.15, 0.2, 0.25, 0.3},
        yticklabel={\pgfmathparse{\tick*100}\pgfmathprintnumber[fixed, zerofill, precision=0]{\pgfmathresult}\%},
        xlabel={Covariance ($2 \times \limfunc{Cov}(\alpha, \psi)$)},
        ylabel={Variance of firm effects ($\limfunc{Var}(\psi)$)},
        axis line style={draw=black!80},
        extra x ticks={0},
        extra x tick style={grid=none, tick style={draw=none}, tick label style={draw=none}},
        extra y ticks={0},
        extra y tick style={grid=none, tick style={draw=none}, tick label style={draw=none}}
        ]
        
        \draw [dashed, thick] (axis cs:0,0) -- (axis cs:0,0.30);
        \draw [dashed, thick] (axis cs:-0.15,0) -- (axis cs:0.20,0);
        
        \tikzset{
            fedot/.style={fill=gray, circle, inner sep=1.2pt},
            targetcircle/.style={draw=blue!70!black, circle, inner sep=2.5pt, thick},
            triarrow/.style={
                -{Triangle[length=5.6pt, width=5pt, gray]},
                densely dotted, 
                black!60, 
                thin,
                shorten >= 0.8pt 
            }
        }
        
        
        \draw[triarrow] (axis cs:0.011, 0.122) -- (axis cs:0.135, 0.055);
        \node[fedot] at (axis cs:0.011, 0.122) {};
        \node[targetcircle] (us_p) at (axis cs:0.135, 0.055) {};
        \node[right=7pt] at (us_p) {US};
        
        \draw[triarrow] (axis cs:0.047, 0.187) -- (axis cs:0.105, 0.153);
        \node[fedot] at (axis cs:0.047, 0.187) {};
        \node[targetcircle] (aus_p) at (axis cs:0.105, 0.153) {};
        \node[right=7pt] at (aus_p) {Austria};
        
        \draw[triarrow] (axis cs:-0.013, 0.231) -- (axis cs:0.087, 0.175);
        \node[fedot] at (axis cs:-0.013, 0.231) {};
        \node[targetcircle] (ita_p) at (axis cs:0.087, 0.175) {};
        \node[right=7pt] at (ita_p) {Italy};
        
        \draw[triarrow] (axis cs:-0.077, 0.244) -- (axis cs:0.113, 0.139);
        \node[fedot] at (axis cs:-0.077, 0.244) {};
        \node[targetcircle] (nor_p) at (axis cs:0.113, 0.139) {};
        \node[right=7pt] at (nor_p) {Norway};
        
        \draw[triarrow] (axis cs:-0.081, 0.146) -- (axis cs:0.039, 0.082);
        \node[fedot] at (axis cs:-0.081, 0.146) {};
        \node[targetcircle] (swe_p) at (axis cs:0.039, 0.082) {};
        \node[right=7pt] at (swe_p) {Sweden};
        
    \end{axis}
\end{tikzpicture}
		\end{center}

	\vspace{-0.2cm}
	
	\begin{quote}
		{\footnotesize
			\textit{Notes:} The figure is constructed from Table F.2 in the online appendix to \citet{bonhomme2023much}. We show the share of variance explained by the variance of firm effects on the y axis, and the share explained by the covariance between worker and firm effects on the x axis. The numbers correspond to 6-year panels, in the largest connected set. We show uncorrected AKM estimates as dots, and bias-corrected estimates from \citet{andrews2008high} in empty circles.}
	\end{quote}	
\end{figure}
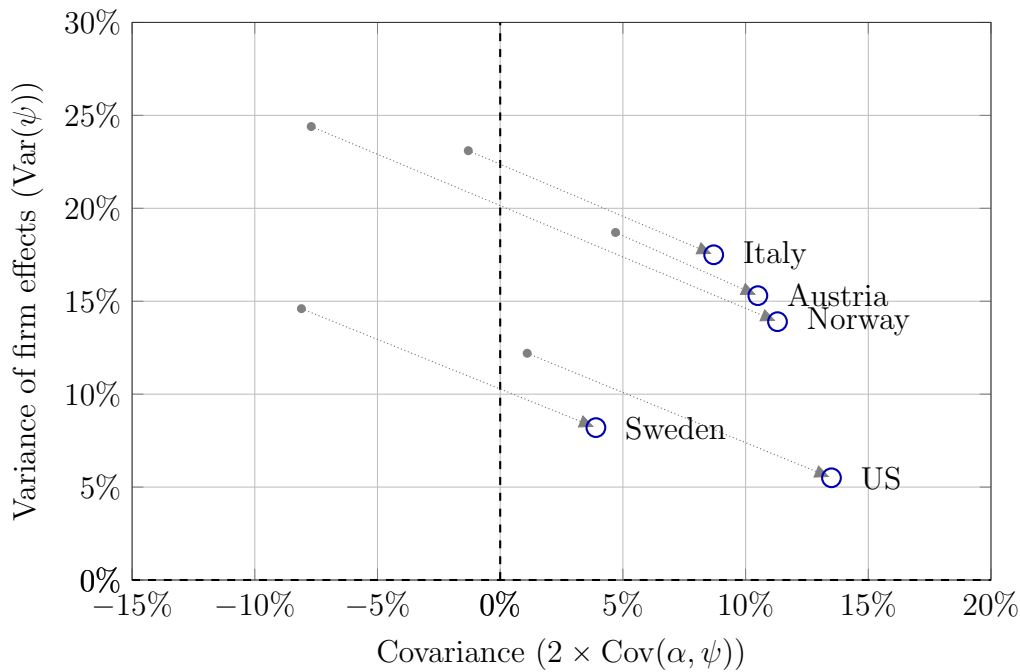

\vskip .3cm \noindent Fixed-effects methods exploit linear regression algebra to construct an estimate of the bias on the quantity of interest, such as the variance of firm effects $\psi_j$. This construction requires specific assumptions on the idiosyncratic shocks. Under the assumption that shocks have a common variance, \citet{andrews2008high} propose an estimator of the variance of $\psi_j$ that is unbiased using a trace-based correction. Their approach equally applies to other variance and covariance terms in standard decompositions. In Figure \ref{fig:literature} we have seen that the correction can materially affect one's conclusions. \citet{kline2020leave} propose an extension of the trace-based correction that allows for unrestricted variance heterogeneity in idiosyncratic shocks.

\vskip .3cm \noindent Correlated random-effects methods rely on a model for the distribution of worker and firm components $\alpha_i$ and $\psi_j$ in the economy. Estimates of the parameters of this distribution can be used to produce estimates of the quantities of interest,  including (but not limited to) variance shares. Moreover, from a Bayesian perspective, one can interpret the distribution as a prior and report posterior average quantities in the spirit of Empirical Bayes. \citet{woodcock2015match} pioneered this approach in the AKM setting, while \citet{bonhomme2023much} and recently \citet{cheng2025optimal} proposed and estimated more flexible models.

\vskip .3cm \noindent To illustrate the magnitude of the bias in practice, in Figure \ref{fig:literature} we plot original estimates based on AKM as well as their bias-corrected counterparts. We report results for five countries, using data from \citet{bonhomme2023much}. We show the bias correction proposed by \citet{andrews2008high}, but other methods (\citealp{kline2020leave}, \citealp{bonhomme2023much}) tend to give similar or larger differences on these data. The figure shows that the original estimates of the variance share of firm effects (on the y axis) range between 12\% and 24\% of the total variance, whereas their bias-corrected counterparts range between 5\% and 17\%. In turn, the AKM estimates of the covariance share (on the x axis) are negative for three of the five countries, whereas the bias-corrected counterparts are all positive, ranging between 4\% and 13\% of the total variance. These findings confirm that, without correction, AKM estimates of variances tend to be overstated while covariances tend to be biased downwards, and they demonstrate that the correction can have a material impact.

\section*{The Frontier of AKM}

\vskip .3cm \noindent The AKM approach introduced by \citet{abowd1999high} has been highly influential in labor economics and other fields. However, the model's assumptions are not uncontroversial, and they are the subject of a growing literature. 

\vskip .3cm \noindent AKM fundamentally relies on two key assumptions: that the log wage depends on the sum of the worker and firm components $\alpha_i+\psi_j$ (\emph{additivity}), and that the idiosyncratic shocks are unrelated to the worker--firm matching process (\emph{exogenous mobility}). When interpreted through the lens of economic models of the labor market, both assumptions are restrictive.

\vskip .3cm \noindent Additivity rules out the presence of interactions $\alpha_i\times \psi_{j}$. However, in many models following the classical theory of sorting proposed by \citet{becker1973theory}, production complementarities between worker and firm inputs are key drivers of sorting patterns. Hence, assuming a particular additive functional form for how worker and firm components affect wages may be restrictive, as emphasized by \citet{eeckhout2011identifying}. 

\vskip .3cm \noindent In turn, exogenous mobility implies that the AKM model cannot account for a worker leaving the firm because of a negative wage shock. Moreover, the model does not allow for the history in past firms to affect a worker's wage. For example, only the new firm affects the wage after a job move, but the model rules out an effect of the previous firm. These features are at odds with search models with wage posting or offer/counteroffer mechanisms (\citealp{postel2002equilibrium}). {They also preclude the presence of time-varying, firm-specific human capital.}   

\vskip .3cm \noindent In reaction to these concerns, researchers have developed a number of diagnostics aimed at assessing the plausibility of the AKM model's assumptions. Notably, \citet{card2013workplace} present an event study graph where they plot wage changes around a job move event. They interpret the graph as suggesting that additivity between worker and firm effects may be a good approximation, and that the absence of a pre-mobility ``dip'' in earnings alleviates concerns about the exogenous mobility assumption.

\vskip .3cm \noindent {However, such diagnostics may not fully resolve concerns about the AKM assumptions.} Researchers have documented empirical violations of additivity, finding that different groups, such as men and women, or racial groups, are affected by different firm effects (\citealp{card2016bargaining}, \citealp{gerard2021assortative}). This suggests that AKM's key assumption that the firm effect $\psi_j$ is the same for all workers does not hold. Moreover, the static nature of AKM is difficult to reconcile with the influential body of work on dynamic economic models of job mobility and wage determination.

\vskip .3cm \noindent The literature is only starting to explore model specifications that account for mechanisms absent in AKM. \citet{bonhomme2019distributional} propose and estimate a wage model that allows for interaction effects between worker and firm heterogeneity. In addition, the framework they introduce allows for dynamic effects of past firms on future wages, and relaxes the exogenous mobility assumption. Their findings based on Swedish data suggest that a log-additive model provides a reasonable approximation to the variance of log wages. At the same time, they find evidence of interactions between worker and firm effects, dynamic effects of past firms, and failure of exogenous mobility, all of which are assumed away in AKM. \citet{abowd2019modeling} propose a model of wages and job flows that explicitly allows for endogenous mobility {by letting the decision to leave the firm be influenced by past idiosyncratic shocks}. 

\vskip .3cm \noindent A parallel development is the growing literature on structural models of the labor market with two-sided -- worker and firm -- heterogeneity. As recent examples, \citet{hagedorn2017identifying} and \citet{sorkin2018ranking} estimate workers' preferences for firms and relate them to AKM firm effects; \citet{card2016bargaining} and \citet{lamadon2022imperfect} bring in information about the value added of the firm and use structural models of job choice and wage determination to study the pass-through of productivity shocks; \citet{lentz2023anatomy} and \citet{lamadon2024labor} estimate dynamic structural models of wages and mobility with two-sided heterogeneity; and \citet{borovivckova2024assortative} focus on the selection issue caused by failure of exogenous mobility within a search model of the labor market.

\vskip .3cm \noindent Methodological developments are also needed for other settings, beyond the traditional labor market applications of the AKM methodology. For example, interaction effects between the $\alpha_i$ of different workers are likely to {be} important in settings with economic spillovers or team production (\citealp{arcidiacono2012estimating}, \citealp{cornelissen2017peer}, \citealp{ahmadpoor2019decoding}, \citealp{bonhomme2021teams}). Likewise, accounting for complementarity between managers' talent and the tasks they perform requires extending the AKM model by allowing for interactions between $\alpha_i$ and $\psi_j$ (\citealp{crippa2025identification}).\footnote{Two recent studies by \citet{mourot2025should} and \citet{villacorta2023unobserved} emphasize the importance of surgeon--hospital and bank--firm interactions, respectively.} Methods to group similar individuals and firms, say, together (\citealp{bonhomme2019distributional}) are useful to reduce the dimensionality of the model and relax various of the key assumptions in AKM. However, there is a need for more theoretical work and applications.

\clearpage

{
	\bibliography{biblio}
}

\end{document}